\documentclass[conference]{IEEEtran}
\IEEEoverridecommandlockouts
\usepackage{cite}
\usepackage{amsmath,amssymb,amsfonts}
\usepackage{algorithmic}
\usepackage{graphicx}
\usepackage{textcomp}
\usepackage{xcolor}
\usepackage{acronym}
\usepackage{gensymb}
\usepackage[caption=false]{subfig}
\usepackage{stfloats}

\acrodef{isac}[ISAC]{Integrated Sensing and Communication}
\acrodef{6g}[6G]{Sixth Generation}
\acrodef{bs}[BS]{Base Station}
\acrodef{5g}[5G]{Fifth Generation}
\acrodef{nlos}[NLoS]{Non Line of Sight}
\acrodef{mmwave}[mmWave]{millimeter Wave}
\acrodef{atr}[ATR]{Automatic Target Recognition}
\acrodef{ml}[ML]{Machine Learning}
\acrodef{dl}[DL]{Deep Learning}
\acrodef{nn}[NN]{Neural Network}
\acrodef{uwb}[UWB]{Ultra WideBand}
\acrodef{ssb}[SSB]{Synchronization Signal Block}
\acrodef{poc}[PoC]{Proof of Concept}
\acrodef{fr2}[FR2]{Frequency Range 2}
\acrodef{gnb}[gNB]{gNodeB}
\acrodef{ru}[RU]{Radio Unit}
\acrodef{ofdm}[OFDM]{Orthogonal Frequency Division Multiplexing}
\acrodef{tdd}[TDD]{Time Division Duplexing}
\acrodef{dft}[DFT]{Discrete Fourier Transform}
\acrodef{idft}[IDFT]{Inverse Discrete Fourier Transform}
\acrodef{nas}[NAS]{Neural Architecture Search}
\acrodef{yolo}[YOLO]{You Look Only Once}

\def\BibTeX{{\rm B\kern-.05em{\sc i\kern-.025em b}\kern-.08em
    T\kern-.1667em\lower.7ex\hbox{E}\kern-.125emX}}
\begin{document}
\bstctlcite{BSTcontrol}

\title{Target Classification for Integrated Sensing and Communication in Industrial Deployments 
}

\author{\IEEEauthorblockN{Luca Barbieri\textit{$^{1}$}, Marcus Henninger\textit{$^{1}$}, Paolo Tosi\textit{$^{1}$}, Artjom Grudnitsky\textit{$^{1}$}, Mattia Brambilla\textit{$^{2}$}, \\ Monica Nicoli\textit{$^{2}$}, Silvio Mandelli\textit{$^{1}$}}
\IEEEauthorblockA{\textit{$^{1}$Nokia Bell Labs, Stuttgart, Germany}, \textit{$^{2}$Politecnico di Milano, Milan, Italy}}

\thanks{This work has received support from the European Commission
through the ISLANDS project (grant agreement no. 101120544).
}
}

\maketitle

\begin{abstract}
Integrated Sensing and Communication (ISAC) systems enable cellular networks to jointly operate as communication technology and sense the environment. 
While opportunities and potential performance have been largely investigated in simulations, few experimental works have showcased Automatic Target Recognition (ATR) effectiveness in a real-world deployment based on cellular radio units.
To bridge this gap, this paper presents an initial study investigating the feasibility of ATR for ISAC.
Our ATR solution uses a Deep Learning (DL)-based detector to infer the target class directly from the radar images generated by the ISAC system. 
The DL detector is evaluated with experimental data from a ISAC testbed based on commercially available mmWave radio units in the ARENA 2036 industrial research campus located in Stuttgart, Germany. 
Experimental results demonstrate accurate classification performance, demonstrating the feasibility of ATR ISAC with cellular hardware in our setup. We finally provide insights about the open generalization challenges, that will fuel future work on the topic.


\end{abstract}


\section{Introduction}
\label{sec:intro}




\ac{6g} cellular networks are transforming the telecommunication industry, accommodating the ever increasing demand for higher data rates, ubiquitous connectivity and ultra high reliability information exchange. 
One technology set to revolutionize this landscape is \ac{isac}~\cite{isac_survey,mandelli_survey}, whereby part of the communication resources is reused for sensing purposes, allowing \acp{bs} to behave as radar devices.
Compared with previous \ac{5g} networks, these radar-like functionalities enable new use cases and applications, such as smart manufacturing~\cite{smart_factory_isac}, vulnerable road user protection~\cite{vru_isac} and intrusion detection~\cite{tosi_nlos}, where the \acp{bs} can assist with complementary information about the possible presence of targets in the environment, ultimately enhancing safety. 

While the integration of \ac{isac} functionalities into \ac{6g} networks provides clear advantages compared with legacy \ac{5g} solutions, target detection based on \ac{isac} brings additional challenges related to the communication-sensing co-design. 
For instance, compared with traditional \ac{mmwave} radars, \ac{isac} systems need to share communication and sensing resources limiting the usability of the resulting radar images~\cite{tosi_nlos,henninger_tdd}.
An example of this are unwanted artifacts due to utilizing standardized \ac{5g} \ac{tdd} frame structures originally not designed for sensing purposes, impacting the final radar image quality.
These hurdles prevent an accurate target identification, calling for dedicated solutions able to adapt to these phenomena and improve object recognition performances.
This is often accomplished by relying on \ac{ml} strategies thanks to their ability of learning complex relationships in a fully data-driven fashion.  
 
Target detection, often referred in the radar jargon as \ac{atr}, has been widely investigated in previous years, especially for \ac{mmwave} radars (see e.g.,~\cite{radar_atr_survey} for a review). 
On the other hand, \ac{atr} based on \ac{isac} systems has been explored only recently.
For example, in~\cite{eye_beam} the authors develop a programmable \ac{isac} platform relying on commercial \ac{5g} hardware and demonstrate the feasibility of \ac{ml}-based \ac{atr} for concealed and non-concealed objects. 
Fall detection is investigated in~\cite{uwb_isac_fall} for \ac{isac}-based \ac{uwb} systems exploiting convolutional \acp{nn}, while~\cite{5g_passive_actvity_rec} focuses on \ac{dl}-based gesture recognition relying on information extracted from \ac{5g} \acp{ssb}.
Other works instead concentrate on WiFi \ac{isac} where the goal is to infer human activity via transformer-based \acp{nn}~\cite{wifi_isac_transformer}, to jointly carry out gesture recognition and human identification by means of 3D convolutional \acp{nn}~\cite{wifi_human_identification} or to localize and count people~\cite{savazzi_counting,savazzi_passive}.
Even though previous works confirm the feasibility of \ac{atr} and similar recognition tasks, e.g. gesture, activity and human identification exploiting \ac{isac} systems, there is a lack of works based on realistic cellular implementations.    
Thus, we aim to bridge this gap by investigating \ac{atr} in \ac{mmwave} \ac{isac} systems based on \ac{5g} commercial hardware. 

In this paper, we aim to investigate the feasibility of \ac{atr} for cellular \ac{isac} systems implemented on real-world hardware.
Our study exploits \ac{dl}-based strategies to infer the target class directly from the radar images generated by the \ac{isac} solution.
The goal is to provide an initial study highlighting key findings and takeaways that apply to cellular \ac{isac} and demonstrate the possibilities of exploiting \ac{isac} for object recognition tasks.   
To do so, we set up an extensive experimental campaign using a \ac{mmwave} \ac{isac} \ac{poc} based on commercially-available \ac{5g} hardware deployed inside the ARENA 20236, an industrial research campus located in Stuttgart, Germany and considering different targets classes, ranging from small objects, i.e., corner reflectors, to large ones, e.g., forklifts. 
Experimental results show that integrating \ac{dl}-based \ac{atr} solutions within cellular \ac{isac} is promising. 
Indeed, we achieve an average accuracy of more than 99\%, indicating that it is possible to infer the correct target class in almost all cases considered.
Furthermore, we also shine light on generalizability issues that arise in \ac{dl}-based \ac{atr}: horizontally flipping the radar images during testing considerably reduces the performances of the \ac{dl} detector for some classes. 
Thus, proper countermeasures are needed to improve performances under unseen testing conditions. 
We leave this extension as a future research activity.


The rest of the paper is organized as follows:
Sec.~\ref{sec:sys_model} introduces the \ac{isac} \ac{poc} and the radar image generation process, while Sec.~\ref{sec:ml_atr} presents the general architecture of the \ac{dl}-based \ac{atr} detector used in the paper. 
Sec.~\ref{sec:meas_setup} describes the experimental activity done for performance assessment, whose results are analyzed in Sec.~\ref{sec:exp_results}.
Finally, Sec.~\ref{sec:conclusions} draws conclusions.



\section{mmWave ISAC real-world setup}
\label{sec:sys_model}


\begin{figure*}
    \centering
    \subfloat[\label{fig:pf_0} $F = 0$]{\includegraphics[width=0.33\linewidth]{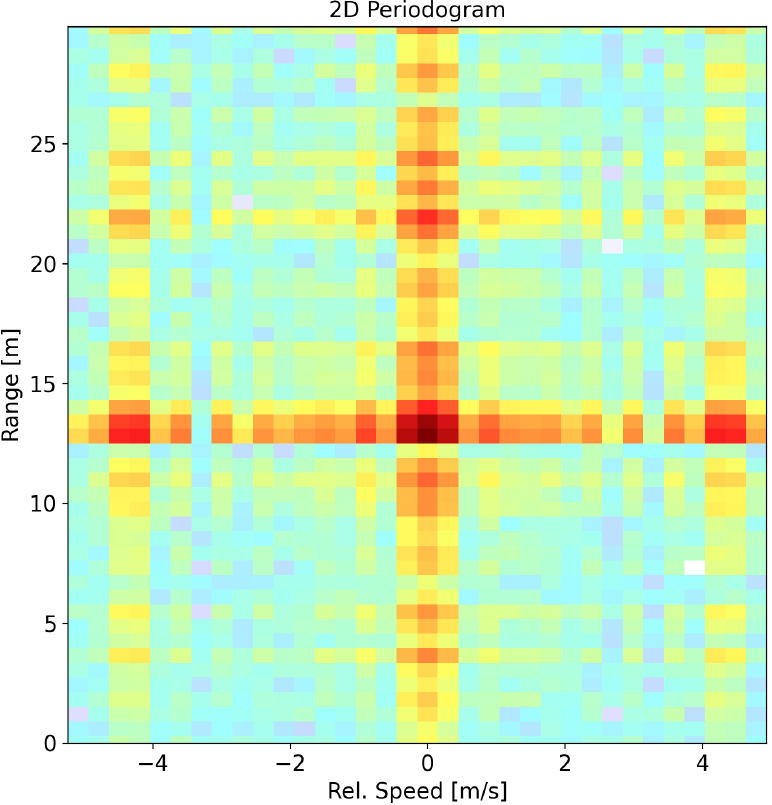}}
    \subfloat[\label{fig:nf_2} $F = 2$]{\includegraphics[width=0.33\linewidth]{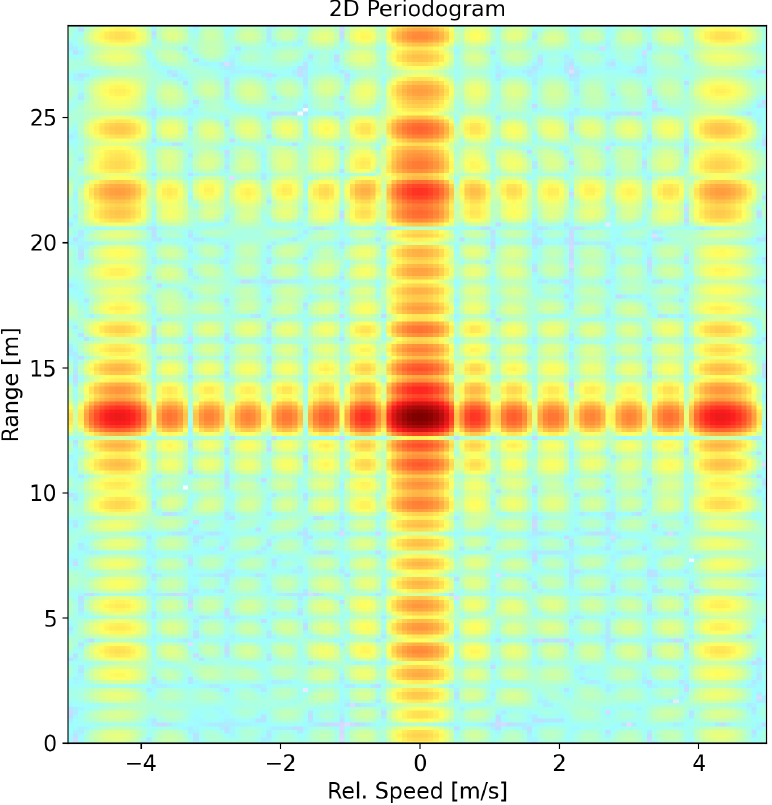}}
    \subfloat[\label{fig:pf_4} $F = 4$]{\includegraphics[width=0.33\linewidth]{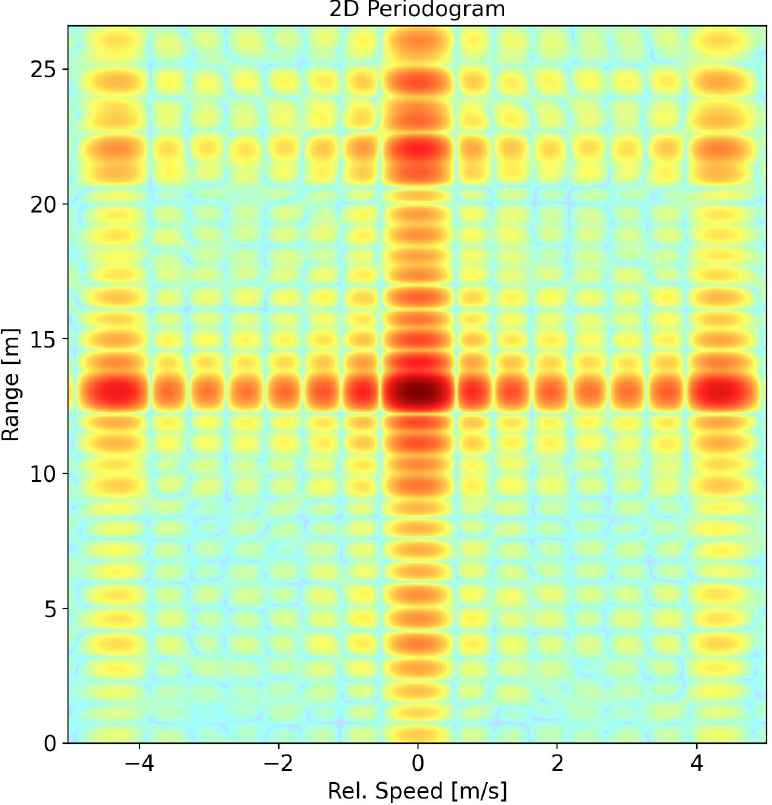}}

    \vspace{-1mm}
    \caption{Effect of different $F$ values on the resulting periodogram $\mathbf{P}$ containing a cabinet target: (a) $F = 0$, (b) $F = 2$, (c) $F = 4$.}
    \label{fig:pf_comparison}
\end{figure*}

This section details the real-world \ac{isac} setup used for collecting the data. Sec. \ref{subsec:icas_poc} presents the \ac{isac} \ac{poc} together with the main parameters characterizing it, while Sec. \ref{subsec:radar_gen} explains how the data extracted from the \ac{poc} is converted to a suitable radar image that can be used for \ac{ml}-based \ac{atr}.

\subsection{mmWave ISAC PoC}
\label{subsec:icas_poc}

\begin{table}[!b]
    \renewcommand{\arraystretch}{1.1}
    \caption{ISAC PoC radio parameters}
    \label{tab:poc_params}
    \centering

    \vspace{-1.5mm}
    \begin{tabular}{|c|c|c|}
    \hline
      \textbf{Parameter} & \textbf{Description} & \textbf{Value} \\ \hline
      $f_c$  & Central frequency & 27.4 GHz\\
      $B$ & Total bandwidth $N \cdot \Delta f$ & 190 MHz\\
      $N$ & Number of subcarriers & 1584 \\
      $M$ & Number of OFDM symbols per frame & 1120 \\
      $\Delta f$ & Subcarrier spacing & 120 kHz\\
      $T_{\text{O}}$ & OFDM symbol time & 8.33 $\mu$s \\
      $T_{\text{CP}}$ & Cyclic Prefix (CP) length & 0.59 $\mu$s \\
      $T_{\text{s}}$ & OFDM symbol time including CP & 8.92 $\mu$s \\ \hline

    \end{tabular}
\end{table}

For our experiments, we rely on the \ac{isac} \ac{poc} described in~\cite{isac_poc} with the main radio parameters summarized in Table \ref{tab:poc_params}. 
This system is implemented using \ac{5g} commercial hardware operating in the \ac{fr2} with a central frequency of 27.4 GHz and consists of two radio devices: a half-duplex \ac{gnb} \ac{ru} configured as the transmitter, and a sniffer \ac{ru} acting as the receiver. 
Note that receiver and transmitter are physically separated but quasi co-located, meaning that the overall system can be regarded as a mono-static sensing setup. 
As far as the actual communication parameters are concerned, 
the \ac{isac} \ac{poc} exploits the \ac{5g} numerology $\mu = 3$~\cite{numerology_3gpp} with $N = 1584$ subcarriers with a spacing among them of $120$ kHz and a total bandwidth of $190$ MHz.  
The \ac{gnb} \ac{ru} transmits in downlink \ac{5g}-compliant \ac{ofdm} radio frames with a duration of $10$ ms using a \ac{tdd} pattern that repeats every $1.25$ ms.
The \ac{tdd} pattern accommodates for $104$ and $36$ symbols for downlink and uplink transmission, respectively, resulting in $M=1120$ total symbols for each radio frame.
On the other hand, the sniffer \ac{ru} operates in uplink and it is synchronized with the \ac{gnb} \ac{ru}. 
Upon reception of the reflected signal by the sniffer \ac{ru}, a dedicated server processes the IQ samples and computes the channel impulse response $\mathbf{H}$ for each radio frame as discussed in the next section.
Note that both \ac{gnb} and sniffer \acp{ru} are configured to use the same fixed beam, with $14 \degree$ half-power beam width (both in azimuth and elevation), for transmitting and receiving the radio signals, respectively. 
For any additional information on the \ac{isac} \ac{poc} we refer the interested reader to~\cite{isac_poc}. 

\subsection{Radar images generation}
\label{subsec:radar_gen}

Let $\mathbf{X} \in \mathbb{C}^{N \times M}$ denote the transmitted \ac{ofdm} frame by the \ac{gnb} \ac{ru} composed by $M$ symbols and $N$ subcarriers. Upon reception of the received signal $\mathbf{Y} \in \mathbb{C}^{N \times M}$ by the sniffer \ac{ru} we can get an estimate of the channel matrix $\mathbf{H} \in \mathbb{C}^{N \times M}$ by element-wise division denoted as 
\begin{equation}
    \mathbf{H} = \mathbf{Y} \oslash \mathbf{X} \, , 
\end{equation}
which carries information about the number of possible targets inside the surrounding environment and their (rough) physical location. 
Given the estimated channel matrix $\mathbf{H}$, we compute the 2D periodogram over the Doppler/delay domain by first applying a \ac{dft} on the \ac{ofdm} symbols and then running an \ac{idft} over the subcarriers.
Each element of the periodogram $\mathbf{P} \in \mathbb{C}^{N' \times M'}$ is expressed as \cite{braun2014ofdm} 
\begin{equation}
    \mathbf{P}_{n,m} = \dfrac{1}{M' N'} \sum_{k=0}^{N'} \left( \sum_{\ell=0}^{M'} \mathbf{H}_{k,\ell} e^{-j 2 \pi \frac{\ell m}{M'}} \right) e^{j 2\pi \frac{k n}{N'}} \, , 
\end{equation}
where $N' = 2^{\lceil{\log_2{N}}  \rceil + F}$ and $M'= 2^{\lceil \log_2{M} \rceil + F}$ are the number of rows and columns of $\mathbf{H}$ after zero padding, while $F$ is the padding factor, a design parameter used for controlling the granularity of the resulting periodogram. 
To get intuitive insights on the effect of different $F$ choices on the generation of $\mathbf{P}$, please see Fig. \ref{fig:pf_comparison}.  
The delay-Doppler information in $\mathbf{P}$ contains all the necessary details to carry out target detection. 

\section{Data-driven ATR system}
\label{sec:ml_atr}

\begin{figure*}
    \centering
    \includegraphics[width=\linewidth]{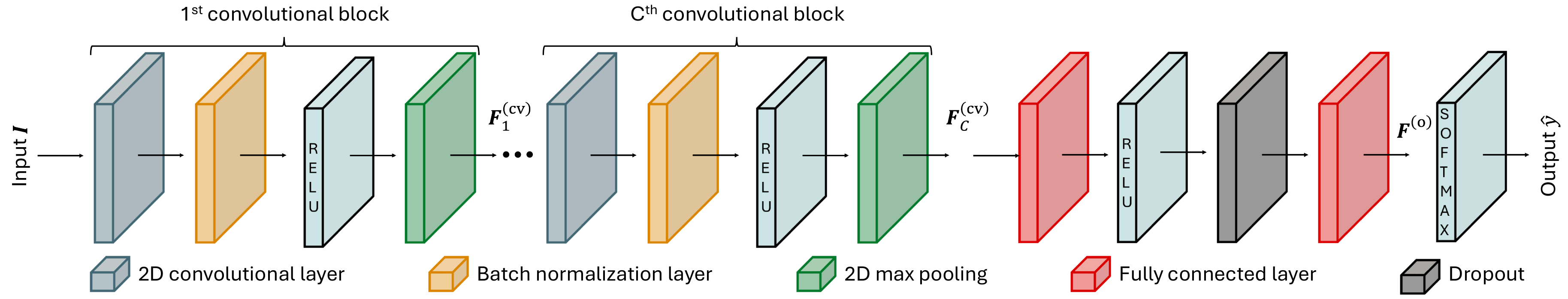}
    
    \vspace{-3mm}
    \caption{Architecture for the convolutional-based \ac{atr} detector: the input is processed by a variable number of convolutional blocks to extract fine-grained features which are fed to two fully connected layers for estimating the final target class.}
    \label{fig:detector}
\end{figure*}





The periodogram generated by the \ac{isac} system is used to understand the presence of targets as well as their physical properties (e.g., discerning different object types).
For this purpose, we rely on \ac{dl} methodologies, namely convolutional \acp{nn}, to infer the type of target being observed by the \ac{isac} system.
Formally, given $\mathbf{P}$ as input, the goal of the \ac{dl}-based detector is to output the object type $\mathbf{y} \in \mathcal{Y}$ among a set $\mathcal{Y}$ of pre-defined target classes with $T = |\mathcal{Y}|$ being the number of possible target classes.
In what follows, we break down the operations performed by the detector, whose general structure is highlighted in Fig. \ref{fig:detector}.

Starting from the complex periodograms, we concatenate magnitude and phase information from $\mathbf{P}$ and feed it to the \ac{dl} model such that each element of the input features $\mathbf{I} \in \mathbb{R}^{N' \times M' \times 2}$ is computed as follows
\begin{equation}
    \mathbf{I}_{m,n} = [\|\mathbf{P}_{m,n} \|, \phi(\mathbf{P}_{m,n})]^{\text{T}} \, ,
\end{equation}
with $\phi(\cdot)$ denoting the phase extraction operation.
Then, we have a variable number $C$ of convolutional blocks whose goal is to extract fine-grained features to help the final classification.
Each block is composed by a 2D convolutional layer followed by a batch normalization layer with a ReLu activation function and a 2D max pooling operation. 
Specifically, the output at the $c$-th block with $c = 1, \ldots, C$ can be generally expressed as 
\begin{equation}
   \mathbf{F}_{c}^{(\text{cv})} = g_{\text{cv}}\left( \mathbf{F}_{c-1}^{(\text{cv})} \right) \, , 
\end{equation}
where $\mathbf{F}_{c-1}^{(\text{cv})}$ are the features of the $(c-1)$-th block with $\mathbf{F}_{c-1}^{(\text{cv})} = \mathbf{I}$ when $c = 1$, while $g_{\text{cv}}(\cdot)$ encapsulates the processing done by the convolutional block.  
After the $C$-th block, we feed $\mathbf{F}_{C}^{(\text{cv})}$ to two fully connected layers with a dropout layer in between to get the final features $\mathbf{F}^{(\text{o})}$for classification.
Specifically, the output of these layers is computed as
\begin{equation}
    \mathbf{F}^{(\text{o})} = g_{\text{fc}}^{(1)}(g_{\text{fc}}^{(2)}(\mathbf{F}_{C}^{(\text{cv})})) \, ,     
\end{equation}
where $g_{\text{fc}}^{(1)}(.)$ and $g_{\text{fc}}^{(2)}(.)$ denote the processing performed by the first and second fully connected layers, respectively. 
The final class is then obtained via a softmax operation $g_{\text{sfx}}(.)$ applied to the output of the last fully connected layer as
\begin{equation}
    \widehat{\mathbf{y}} = g_{\text{sfx}}(\mathbf{F}^{(\text{o})}) \, . 
\end{equation}
The \ac{dl} model parameters are optimized by minimizing a weighted version of the cross entropy loss between the predicted label $\hat{\mathbf{y}}$ and the true one $\mathbf{y}$ for the samples in the training set. 
This is done to account for possible imbalances, i.e., number of training examples, across different target classes. 
Note that the final choice of the number of convolutional blocks $C$ as well as the parameters characterizing all the layers of the detector is optimized as detailed in Sec. \ref{sec:exp_results}. 


\section{Measurement campaign}
\label{sec:meas_setup}

\begin{figure*}
    \centering
    \subfloat[\label{fig:class_0}]{\includegraphics[width=0.25\linewidth]{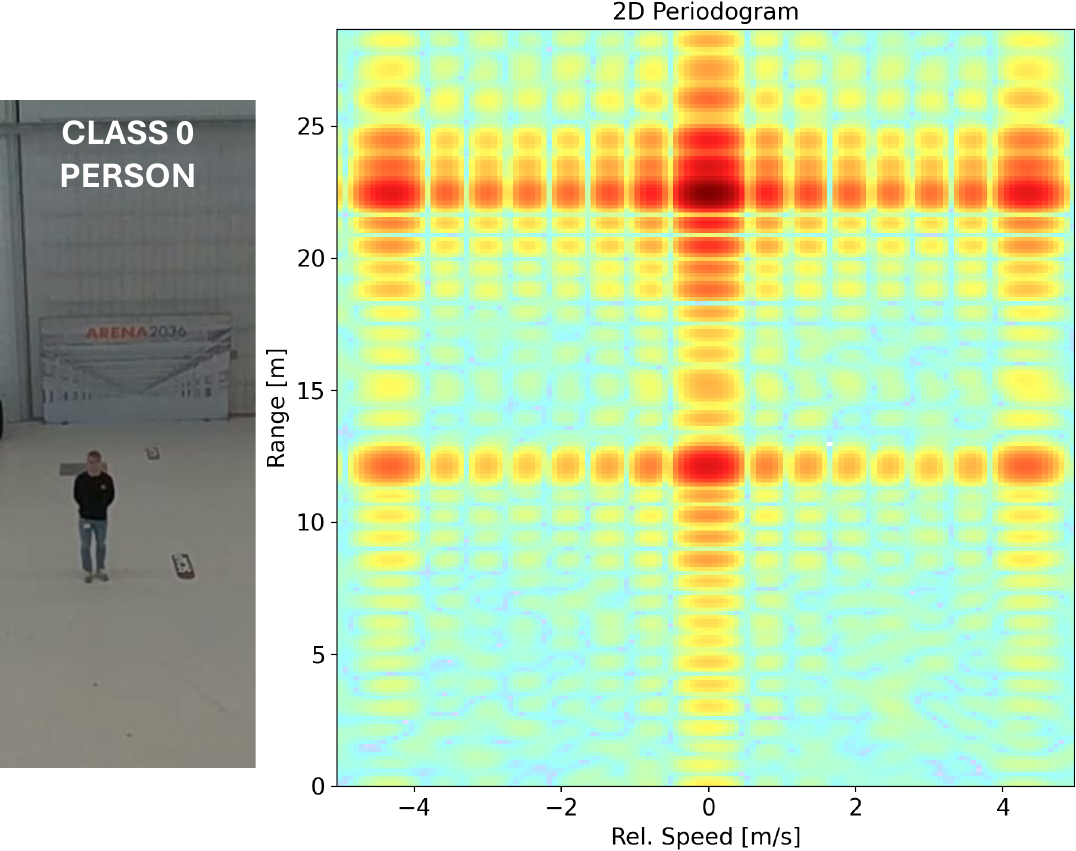}}
    \subfloat[\label{fig:class_1}]{\includegraphics[width=0.25\linewidth]{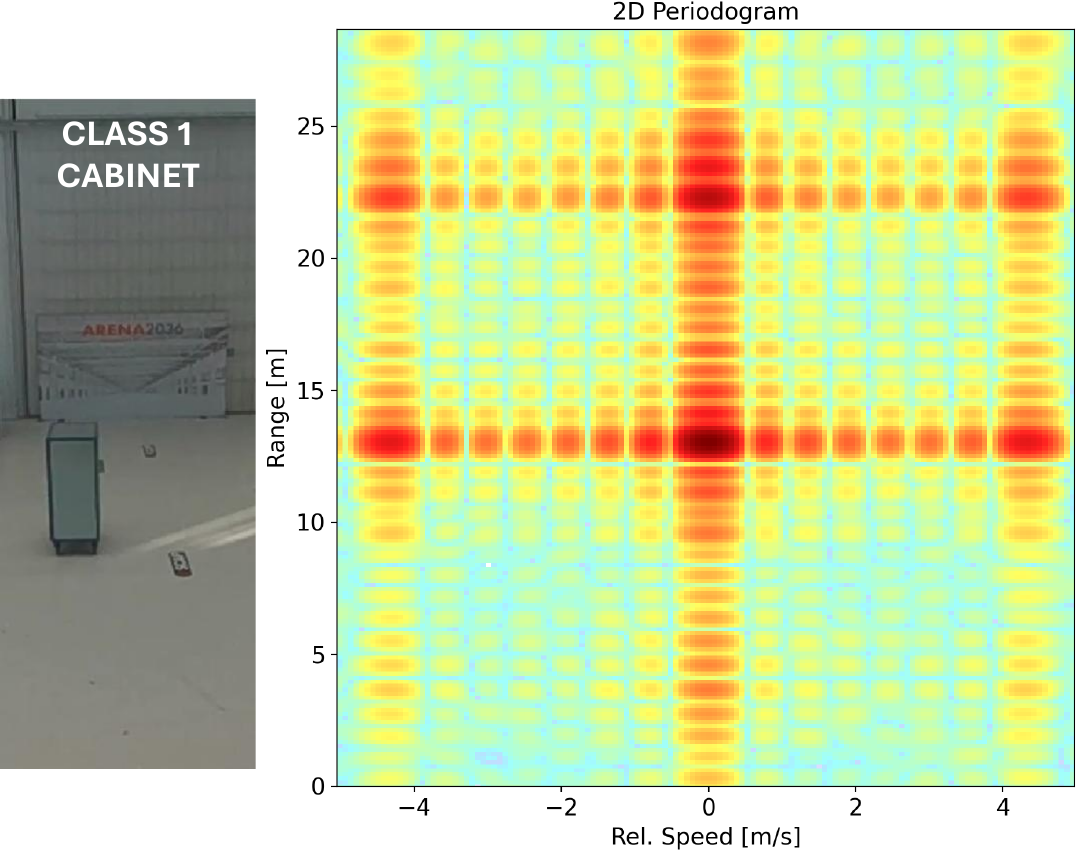}}
    \subfloat[\label{fig:class_2}]{\includegraphics[width=0.25\linewidth]{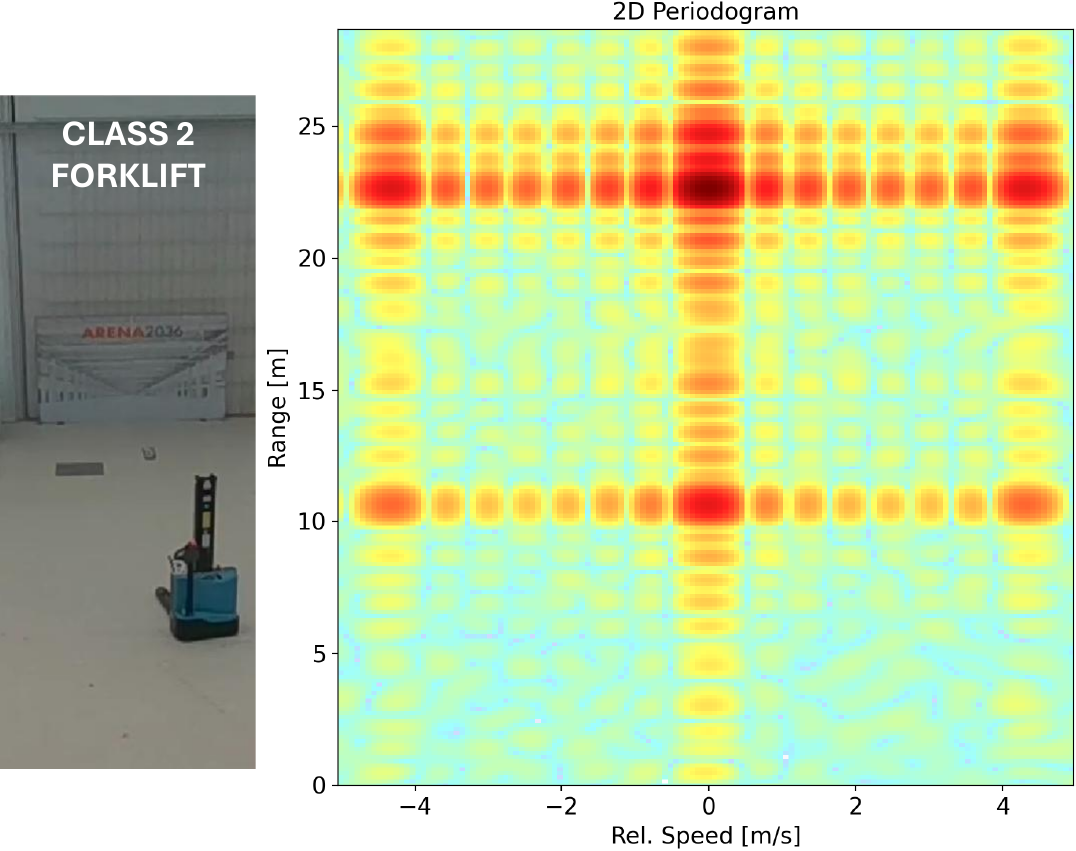}}
    \subfloat[\label{fig:class_3}]{\includegraphics[width=0.25\linewidth]{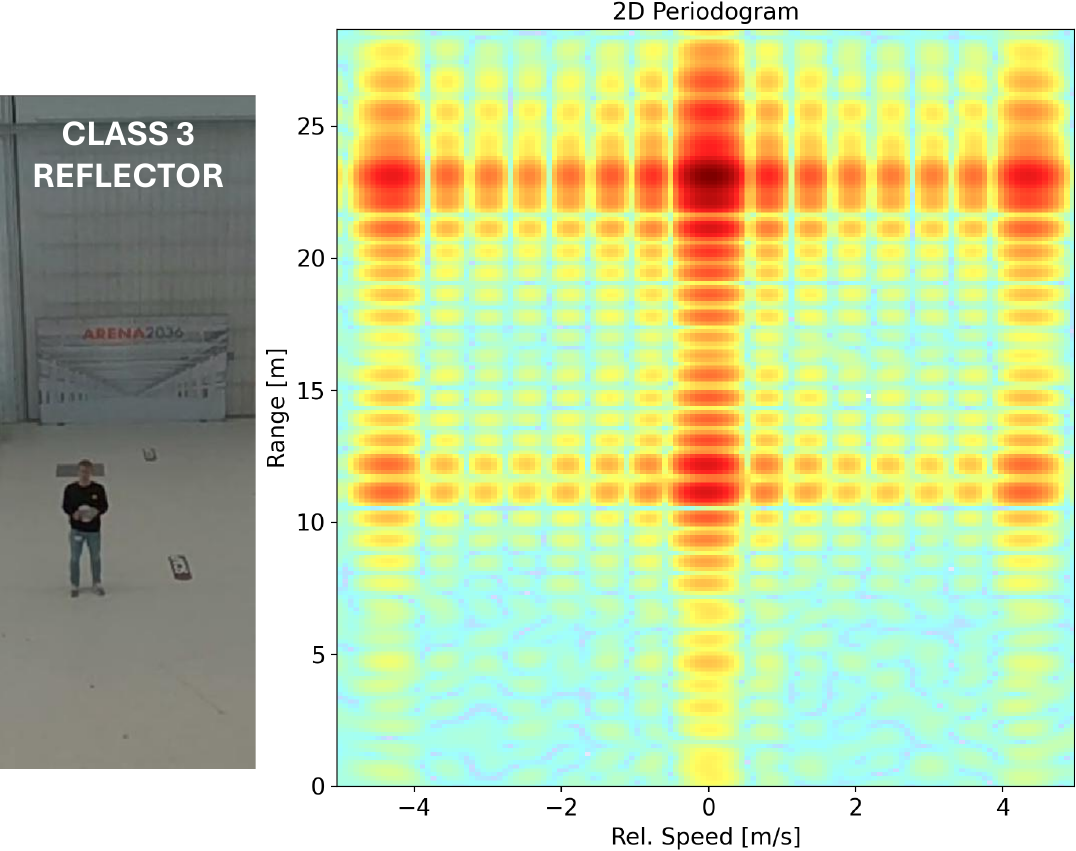}}

    \vspace{-2mm}
    \subfloat[\label{fig:class_4}]{\includegraphics[width=0.25\linewidth]{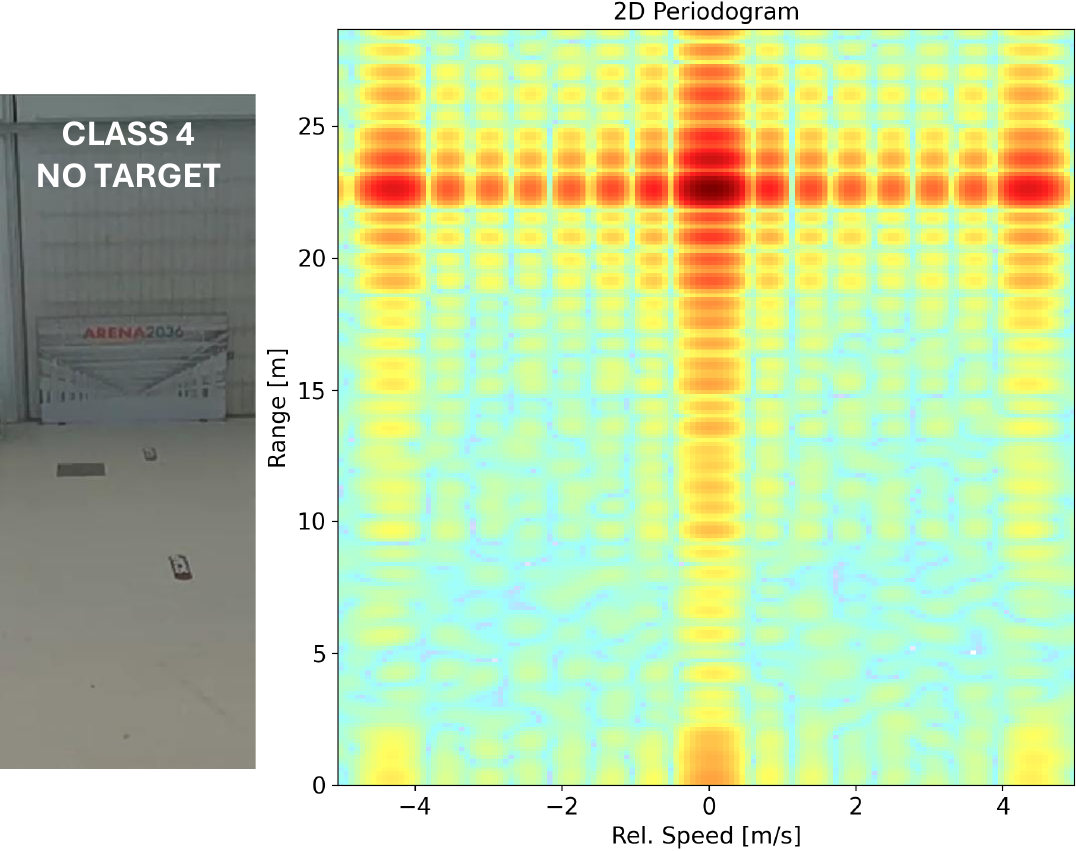}}
    \subfloat[\label{fig:class_5}]{\includegraphics[width=0.25\linewidth]{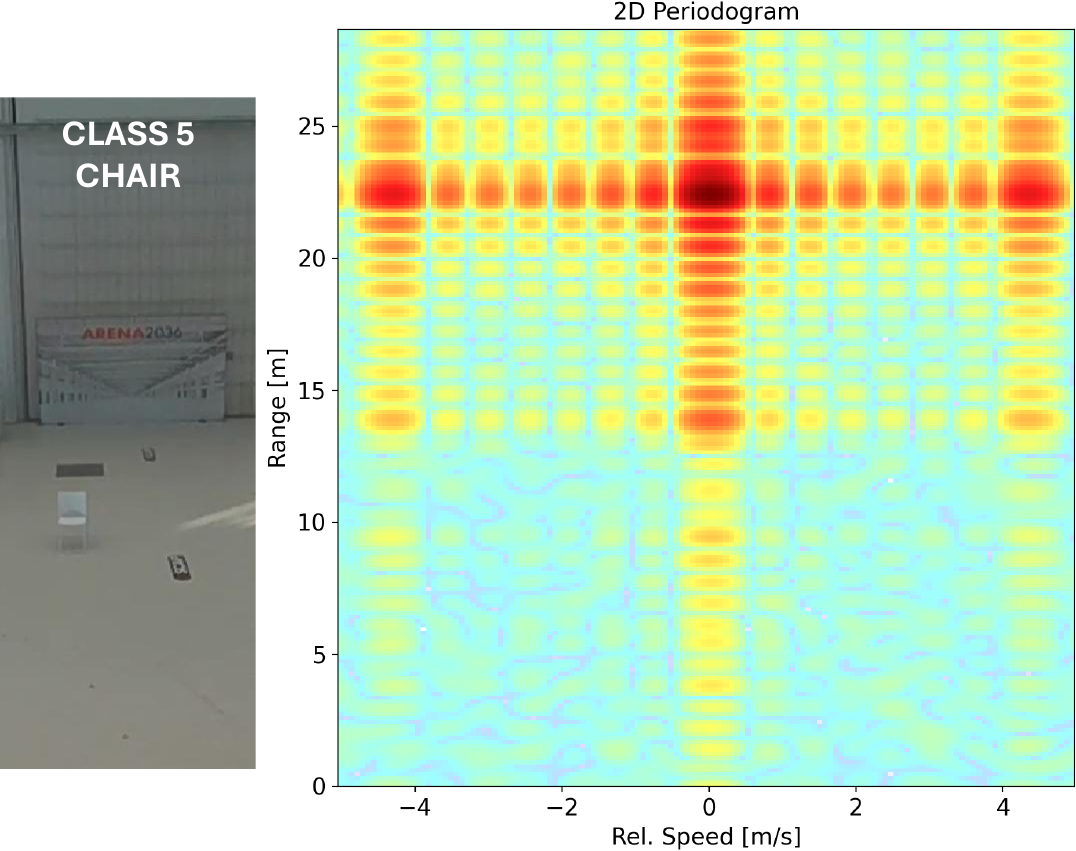}}
    \subfloat[\label{fig:class_6}]{\includegraphics[width=0.25\linewidth]{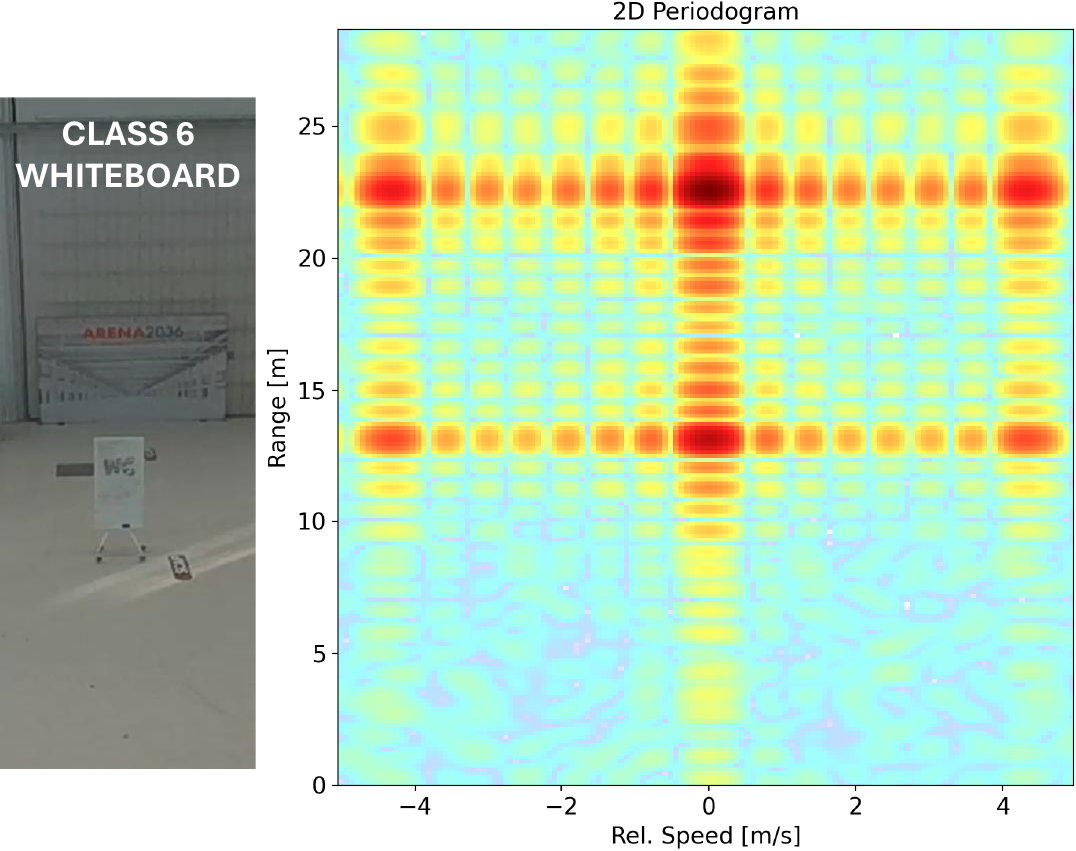}}
    \subfloat[\label{fig:class_7}]{\includegraphics[width=0.25\linewidth]{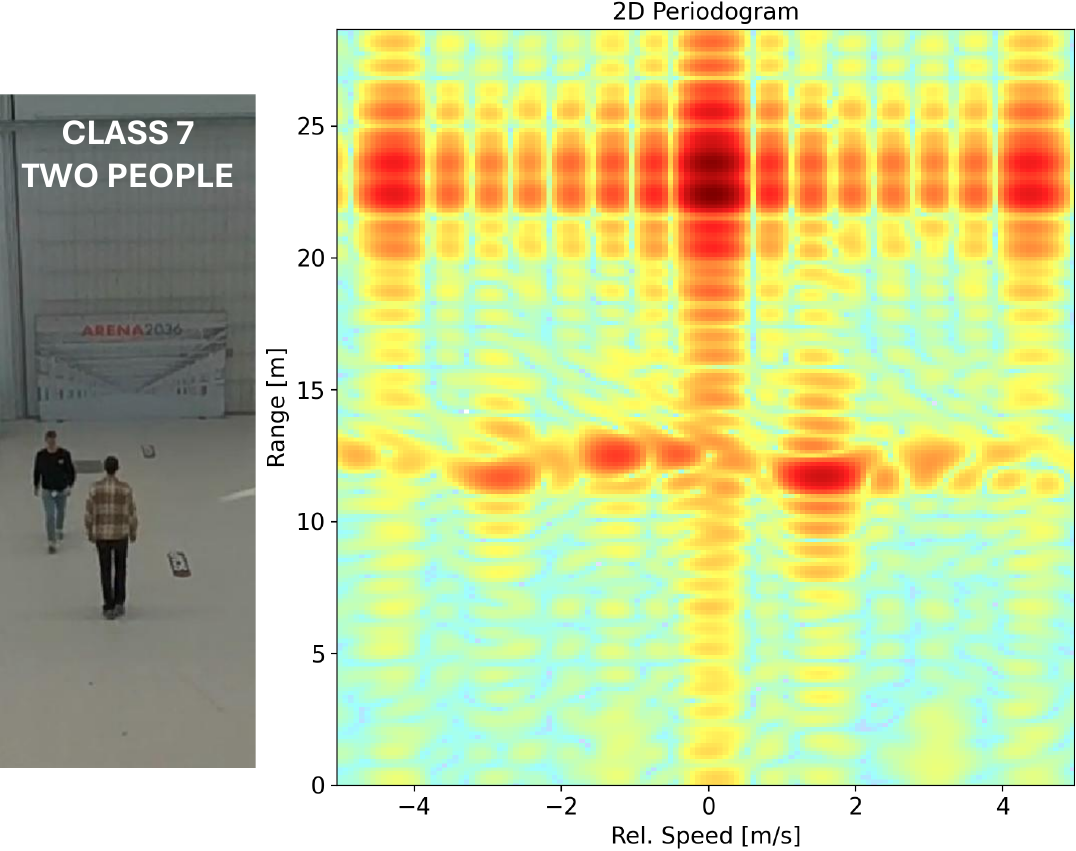}}
    \vspace{-1.5mm}
    
    \caption{Measurements collected during the experimental campaign. Each panel shows the periodogram, on the right, and the corresponding visual representation collected via a camera, on the left, for the targets: (a) person, (b) cabinet, (c) forklift, (d) reflector, (e) no target, (f) chair, (g) whiteboard and (h) two people.}
    \label{fig:dataset_examples}

\end{figure*}

To assess the \ac{atr} feasibility for cellular \ac{isac} systems, we design an extensive experimental campaign aimed at evaluating the accuracy provided by the \ac{dl}-based \ac{atr} detector in correctly classifying a wide variety of different target classes.   

The measurement campaign has been carried out in the ARENA 2036 industrial research campus located in Stuttgart, Germany, where the \ac{isac} \ac{poc} has been mounted at a height of $5.12$ m to surveil an area in front of it. 
The area covers a maximum distance of $23$ m, dictated by a wall with a cargo gate present at the end of the area, and is free from major obstructions, allowing us to capture the effect of each target on the resulting periodograms collected during the campaign.
The experiments focus on collecting \ac{isac} data to form a dataset for \ac{ml} purposes. 
The following target classes and movements have been considered:
\begin{itemize}
    \item Person (class 0): a person is placed at two different distances, approximately at 11 m and 18 m. Then, the person walks in a straight line moving away from the \ac{poc} and back. The same movement is then recorded when the person runs instead of walking. 
    \item Cabinet (class 1): a large metallic cabinet structure is first placed at a fixed distance (roughly 15 m from the \ac{poc}). \textcolor{black}{Then, we move the cabinet in a straight line following the same trajectory for class 0.} 
    \item Forklift (class 2): a small forklift is placed at a fixed distance of 11 m from the \ac{poc}.
    \item Reflector (class 3): \textcolor{black}{a person carrying a small corner reflector is placed at 11 m and 18 m from the \ac{poc}.} As done before, we record also the movement of the person walking/running away from the \ac{poc} and back.   
    \item No target (class 4): in this case the area in front of the \ac{poc} does not contain any target of interest and is free from any obsctruction. 
    \item Chair (class 5): we place a white chair at a fixed distance of 15 m from the \ac{poc}. 
    \item Whiteboard (class 6): similar for class 5, we place a whiteboard at a fixed distance of 15 m from the \ac{poc}.
    \item Two people (class 7): two people move in opposing motion. The first one starts near the \ac{poc} while the second begins close to the end of the area. The velocity of each person varies over time, with parts of the trajectory where they walk and parts of it where they run. 
\end{itemize}

\begin{table}[!b]
    \renewcommand{\arraystretch}{1.1}
    \caption{Measurement campaign statistics}
    \label{tab:dataset_statistics}
    \centering

    \vspace{-1.5mm}
    \begin{tabular}{|c|c|c|c|}
    \hline
    \textbf{Class}  & \textbf{Periodograms} [\#] & \textbf{Ratio} [\%]& \textbf{Train/test samples} [\#] \\ \hline
     0 & 2564 & 19.58 & 2051/513 \\
     1 & 1635 & 12.48 & 1308/327 \\ 
     2 & 1051 & 8.03 & 841/210 \\
     3 & 3345 & 25.55 & 2776/569 \\
     4 & 1108 & 8.46 & 886/222 \\
     5 & 803 & 6.14 & 643/160 \\
     6 & 754 & 5.77 & 604/150 \\
     7 & 1832 & 13.99 & 1466/366 \\ \hline
     Total & 13092 & 100 & 10575/2517 \\ \hline
    \end{tabular}
\end{table}

According to the above discussion, we strive to maximize diversity while collecting data by considering different distances as well as movement types (i.e., walking and running) for most of the targets considered.
This allows us to benchmark the performances of the \ac{dl}-based detector under conditions that are as general as possible and provide more comprehensive conclusions and takeaways.
For each of the considered targets in the dataset, we show in Fig. \ref{fig:dataset_examples} an exemplary periodogram with the associated target captured via a camera system. 
By comparing the different periodograms, some classes show quite similar visual properties that make them hardly distinguishable, calling for dedicated \ac{ml}-based processing techniques. 
Finally, Table \ref{tab:dataset_statistics} highlights the number of periodograms collected for each class, the ratio of data per each class with respect to the total number of periodograms, as well as the number of train/test examples used for \ac{dl}-based \ac{atr} training and testing.


\section{Experimental results}
\label{sec:exp_results}

\begin{figure*}
    \centering
    \subfloat[\label{fig:result_nf0}]{\includegraphics[width=0.33\linewidth]{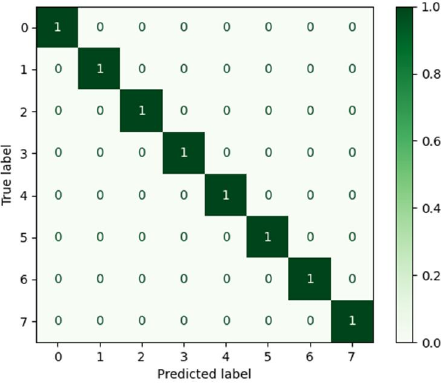}}
    \subfloat[\label{fig:result_nf1}]{\includegraphics[width=0.33\linewidth]{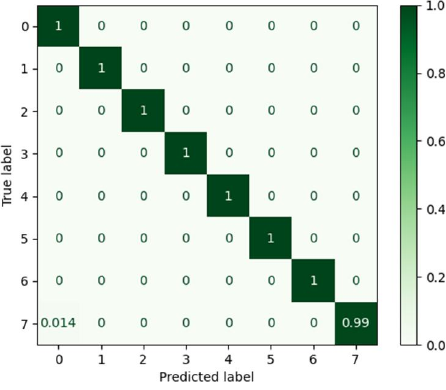}}
    \subfloat[\label{fig:result_nf2}]{\includegraphics[width=0.33\linewidth]{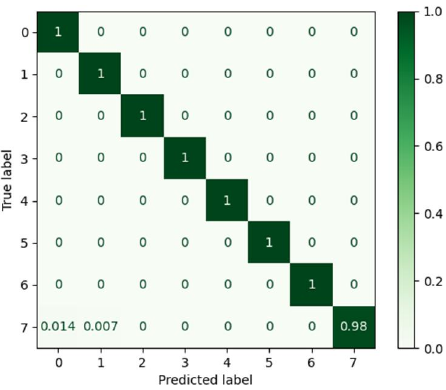}}

    \vspace{-1mm}
    \caption{\textcolor{black}{Confusion matrices attained by the \ac{atr} detector for: (a) $F = 0$, (b) $F = 1$ and (c) $F = 2$. The average accuracy for $F = 0$, $F = 1$ and $F = 2$ is 99.96\%, 99.91\% and 99.91\%, respectively.}
    }
    \label{fig:accuracy_results}
\end{figure*}

This section details the experimental results provided by the \ac{dl}-based \ac{atr} detector. 
We start by describing the optimization process used to find the best architecture in terms of layer choices as well as parameters characterizing each \ac{nn} layer in Sec. \ref{subsec:nn_optimization}.
Then, in Sec. \ref{subsec:atr_results}, we present the results provided by the optimized configuration of the detector, highlighting also possible generalization abilities.

\subsection{NN optimization and best architecture search}
\label{subsec:nn_optimization}

We implement an architecture search on the \ac{nn} model defined in Sec. \ref{sec:ml_atr} to optimize its structure, e.g., number of layers and their associated configuration parameters, according to the value of the padding factor $F$ selected.

We optimize the number $C \in [2,4]$ of convolutional blocks, while for the convolutional layers we search for the best kernel size $k_c \in [7, 5, 3]$, stride size $s_c \in [2, 1]$ and number of output channels $o_c \in [16, 8, 4]$. 
Similarly, we search for the optimal kernel size $k_m \in [2, 1]$ and stride size $s_m \in [2, 1]$ for the max pooling operations performed inside the convolutional blocks. 
We follow a widely adopted choice in the literature of doubling the number of channel outputs after every convolutional layer, see e.g., \cite{vgg} and \cite{resnet}.
In our search $o_c$ refers to the number of output channels of the first layer, while subsequent layers adopt multiple of 2s for the number of output channels. 
As an example of this strategy, if $C=4$ and $o_c=4$, the first convolutional layer has $4$ output channels, the second $8$, the third $16$ and the last one $32$.
Additionally, we fix $k_c$ only of the first convolutional layer according to the $F$ value selected, namely if $F = 0$, $F = 1$ and $F = 2$ we select $k_c = 3$, $k_c = 5$ and $k_c = 7$, respectively. 
Finally, for what concerns the fully connected layers, we search for the best number of output features $f \in [16, 32, 64]$, while also optimizing the dropout rate $d$ with $d \in [0.8, 0.5]$. 
All other parameters are fixed, namely we choose a batch size of $32$ examples, adopt the SGD optimizer with a learning rate of $0.001$ and a  number of epochs for training of $50$ without any learning rate scheduler.  

\begin{table}[!b]
    \renewcommand{\arraystretch}{1.1}
    \caption{Optimized architectures for the chosen values of $F$}
    \label{tab:nn_architectures}

    \vspace{-2mm}
    \centering
    \begin{tabular}{|c|c|c|c|c|c|}
    \hline
    \multicolumn{2}{|c|}{$F = 0$} & \multicolumn{2}{c|}{$F = 1$} & 
    \multicolumn{2}{c|}{$F = 2$} \\ \hline 
    \textbf{Parameter} & \textbf{Value} & \textbf{Parameter} & \textbf{Value} & \textbf{Parameter} & \textbf{Value} \\ \hline
    $C$    & 2   & $C$ & 2 & $C$ & 4  \\ 
    $k_c$  & 3   & $k_c$ & 3 & $k_c$ & 3  \\ 
    $s_c$  & 1   & $s_c$ & 1 & $s_c$ & 1 \\ 
    $o_c$  & 16  & $o_c$ & 16 & $o_c$ & 8\\
    $k_m$  & 2   & $k_m$ & 2 & $k_m$ & 2 \\ 
    $s_m$  & 2   & $s_m$ & 2 & $s_m$ & 2 \\
    $f$    & 32  & $f$ & 32 & $f$ & 32\\
    $d$    & 0.5 & $d$ & 0.5 & $d$ & 0.5 \\ \hline 
    $D$ & 178k & $D$ & 343k & $D$ & 136k \\ \hline

    \end{tabular}
\end{table}

The best architectural design of the \ac{dl}-based \ac{atr} detector is done by randomly configuring the \ac{nn} with the parameters defined above and monitoring the testing loss at the end of the training process. 
We make $80$ different configurations for each of the $F$ values selected, namely $F = 0$, $F = 1$ and $F = 2$, and choose the resulting architecture giving the lowest testing loss for each one of them. 
This process provides at the output the \ac{nn} architectures whose configurations parameters are defined in Table \ref{tab:nn_architectures} together with their number $D$ of trainable parameters. 
Overall, smaller padding factors (i.e., $F = 0$ or $F = 1$) call for a lower number of convolutional blocks for maximizing the performances, while for larger ones (i.e, $F = 2$) one should adopt more convolutional blocks. 
This is reasonable as $F$ controls the granularity of the periodograms: increasing $F$ leads to more fine-grain details in $\mathbf{P}$ ultimately requiring more convolutional layers to process them and fully exploit for the final classification. 


\subsection{ATR accuracy and generalization capabilities}
\label{subsec:atr_results}

As a first set of results, we present the classification accuracy attained by the \ac{dl}-based \ac{atr} detector over the different values of the padding factor $F$. 
Specifically, Fig. \ref{fig:accuracy_results} shows the confusion matrices that highlight the accuracy for each target class individually considering $F = 0$ (Fig. \ref{fig:result_nf0}), $F = 1$ (Fig. \ref{fig:result_nf1}) and $F = 2$ (Fig. \ref{fig:result_nf2}). 
\textcolor{black}{The results show that the performances attained by the ATR detector are not influenced much by the padding factor selected} 
Indeed, the average accuracy attained across the $F$ values is nearly identical, apart from minor differences for the last target class, i.e., the two people case. 
The results thus suggest that exploiting \ac{dl} methodologies for \ac{atr} from \ac{isac} system is highly effective as it provides excellent recognition abilities. 

\begin{figure}[!t]
    \centering
    \includegraphics[width=\linewidth]{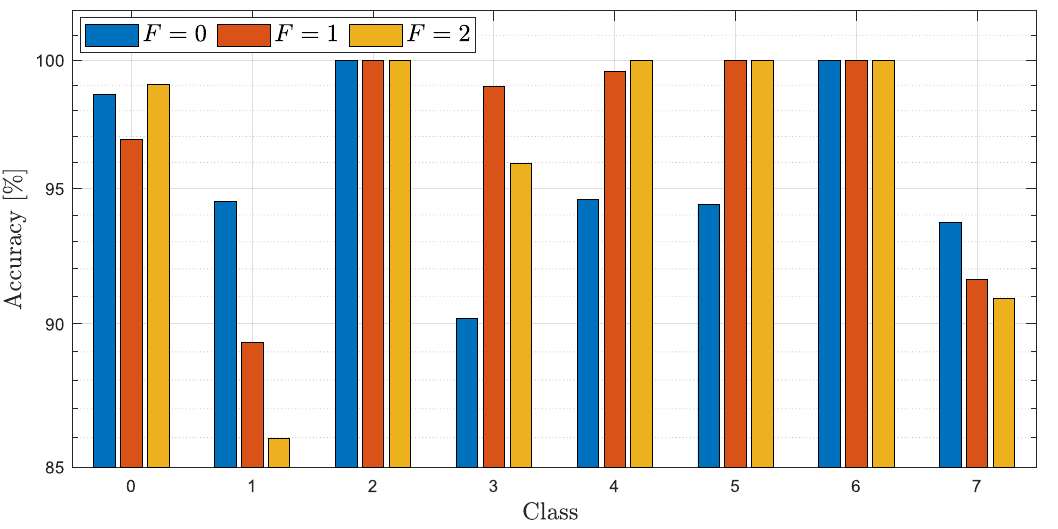}

    \vspace{-4mm}
    \caption{Per-target classification accuracy after horizontal flip for $F = {0, 1, 2}$.}
    \label{fig:nn_generalizability}

\end{figure}

\textcolor{black}{We also evaluate the generalizability of the trained DL
detectors by horizontally shifting all periodogram samples contained in the testing dataset and evaluating its effects on the final classification accuracy} 
Conceptually, this processing does not affect the appearance of the targets or their distance from the \ac{isac} \ac{poc} but merely flips the relative velocity information.
This makes targets moving in a specific direction appear to be moving in the opposite. 
Fig. \ref{fig:nn_generalizability} shows the results of the procedure, highlighting the accuracy per target class attained by the \ac{dl} detectors over the different values of $F$.
Comparing the results, selecting a single $F$ parameter might not be feasible as there is no value guaranteeing the highest performance across all target classes. 
The analysis also suggests that \ac{dl} methods for \ac{atr} could benefit from augmentation strategies or dedicated countermeasures to improve performances in unseen scenarios.
\textcolor{black}{Indeed, under all F values,
the accuracy for some target classes, i.e., person and two people classes, diminishes quite considerably} 
Conventional augmentation strategies currently developed for images do not reflect well the complex interactions captured by the communication channel, calling for ad-hoc augmentations that provide physically plausible transformations to the targets being currently detected by the \ac{isac} system.
We leave this problem for future works.

\section{Conclusions}
\label{sec:conclusions}

In this paper we investigated the feasibility of conducting \ac{atr} fully relying on cellular \ac{isac} systems implemented over real-world \ac{5g} \ac{mmwave} hardware. 
We conducted an extensive experimental campaign comprising 8 different target classes together with varying movement patterns.
Based on the collected data from the \ac{isac} \ac{poc}, we optimized a \ac{dl}-based \ac{atr} detector in terms of architectural choices to maximize its classification accuracy over varying padding factors that rule the granularity of the gathered radar images. 
Experimental results show the feasibility of exploiting \ac{dl} methodologies for \ac{isac} \ac{atr} as they are able to correctly recognize all target classes in nearly all conditions considered.
Nevertheless, introducing data augmentation (i.e, horizontally flipping the periodograms) during testing substantially reduces classification accuracy for some target classes. 
This highlights the need for a generalizable \ac{atr} solution able to adapt to different sensing operations, environments and unseen conditions. 
We strive to tackle this problem in future works by  extending this preliminary study to more complex scenarios by considering a wider range of possible targets and movement combinations, as well as explore dedicated strategies to deal with changing environments and \ac{isac} \ac{poc} parameters configuration, without requiring a full re-training of the \ac{dl}-based \ac{atr} solutions.

\bibliographystyle{IEEEtran}
\bibliography{IEEEabrv,biblio}

\end{document}